\documentclass[10pt,a4paper]{article}
\usepackage{amsmath}
\usepackage{amsfonts}
\usepackage{amssymb}
\usepackage{graphicx}
\usepackage{textcomp}
\usepackage{multicol}
\usepackage{url}
\usepackage{lipsum}
\usepackage{setspace}
\usepackage{comment}
\usepackage{color}
\usepackage[numbers,sort&compress]{natbib}
\usepackage[usenames,dvipsnames,svgnames,table]{xcolor}
\usepackage[lofdepth,lotdepth,caption=false]{subfig}
\usepackage[pdfstartview=FitH]{hyperref}
\usepackage{hyperref}
\newsavebox{\bigleftbox}
\hypersetup{
 colorlinks,
 citecolor=Blue,
 linkcolor=Blue,
 urlcolor=Blue
}

\makeatletter
 \def\footnoterule{\kern-3\p@
   \noindent\hrulefill \kern 2.8\p@} % the \hrule is .4pt high
\makeatother
\usepackage[left=3.00cm, right=3.00cm, top=2.00cm, bottom=2.00cm]{geometry}

\title{\textbf{First-Principles and Machine Learning Insights into the Design of DOTT-Carbon and its Lithium-Ion Storage Capacity}}
\author{
   Kleuton. A. L. Lima$^{1,2}$, 
   Ana V. P. Abreu $^{1,2}$, 
   Alysson M. A. Silva $^{3}$, and \\
   Luiz A. Ribeiro Jr$^{1,2,\S}$
	}
\date{}

\begin{document}
    \maketitle
	\vspace{-0.6cm}
	\begin{center}\small
     \textit{$^{1}$Institute of Physics, University of Bras\'ilia, 70910-900, Bras\'ilia, Brazil}\\
     \textit{$^{2}$Computational Materials Laboratory, LCCMat, Institute of Physics, University of Bras\'ilia, 70910-900, Bras\'ilia, Brazil}\\
     \textit{$^{3}$University of Bras\'ilia, College of Technology, Department of Mechanical Engineering, Bras\'ilia, Brazil}\\
    \phantom{.}\\ \hfill
        $^{\S}$\url{ribeirojr@unb.br}\hfill
		\phantom{.}
	\end{center}
	
% ---------------------- %
%\blfootnote{}
%\blfootnote{}

\onehalfspace

\noindent\textbf{Abstract: Two-dimensional (2D) carbon-based materials are promising candidates for developing more efficient green energy conversion and storage technologies. This study presents a new 2D carbon allotrope, DOTT-Carbon, characterized by its distinctive and multi-ring structure featuring 12-, 8-, 4-, and 3-membered rings of carbon atoms. We explore its structural, mechanical, and lithium-ion storage properties by employing density functional theory and machine learning simulations. Phonon calculations confirm its structural stability and ab initio molecular dynamics simulations demonstrate its thermal resilience at elevated temperatures. The material exhibits anisotropic mechanical properties, with Young's modulus values varying between 280-330 GPa. DOTT-Carbon displays a lithium-ion storage capacity of 446.28 mAh/g, complemented by a low diffusion barrier (0.2-0.9 eV) and a high diffusion coefficient ($ > 1.0 \times 10^{-6}$ cm$^{2}$/s), possibly facilitating efficient lithium-ion transport. The stable open circuit voltage of 0.28 V also indicates its suitability as an anode material.}

\section{Introduction}

Two-dimensional (2D) carbon materials have attracted significant attention due to their unique structural, thermal, and electronic properties \cite{xu2013graphene,balandin2011thermal}. Most have been proven promising candidates for more efficient green energy conversion and storage technologies \cite{tao2020two,kumar2018recent}, particularly for lithium-ion batteries (LIBs) \cite{peng2016two}. Once the demand for sustainable energy storage solutions grows, these materials provide a versatile and cost-effective platform for developing new architectures and improving battery efficiency.

Porous 2D carbon structures have been shown to enhance lithium-ion storage performance \cite{liu2017porous,zheng2015two,fang2013two}. The possibility of forming porous rings comprising more than eight atoms and large surface areas promotes efficient ion adsorption \cite{santos2024proposing,santos2024photh,wang2018popgraphene,al2022two,pasanaje2024evolutionary,li2021two,liu2024novel}. It facilitates rapid diffusion pathways for lithium ions. Graphyne (comprising a 12-atom ring) \cite{desyatkin2022scalable} and the biphenylene network (comprising an 8-atom ring)\cite{fan2021biphenylene}, for example, have demonstrated good lithium-ion storage capacities and fast charge/discharge cycles \cite{yang2019mechanochemical,ferguson2017biphenylene}. These advancements highlight the critical role of porosity in tailoring the properties of 2D carbon-based materials for energy storage applications.

Despite these successful cases, the ongoing search for new 2D carbon allotropes with optimized electrochemical properties remains crucial to advancing nanoelectronics \cite{kharissova2019carbon,tiwari2016magical}. Recent studies emphasize the potential of designing multi-ringed porous architectures to balance mechanical stability and high lithium-ion storage capacity \cite{rajkamal2019carbon,lherbier2018lithiation,wang2021carbon,ali2023advancements,yu2013graphenylene,ruby2024recent,gomez2024tpdh,cai2023lc,lu2022new,cai2021net,cheng2021two,you20242d,liu2017new,wang2022thgraphene,liu2022applications}. In this context, we introduce DOTT-Carbon (DOTT-C), a new 2D all-carbon allotrope characterized by 12-, 8-, 4-, and 3-membered rings, opening channels for a key structural framework for energy storage applications.

Herein, we investigate the structural, mechanical, and lithium-ion adsorption properties of DOTT-C using density functional theory (DFT) calculations and machine learning interatomic potentials (MLIPs). We confirm its stability through phonon calculations and ab initio molecular dynamics (AIMD) simulations. Importantly, the lithium diffusion barriers, storage capacity, and open-circuit voltage of DOTT-C were also explored. This newly proposed 2D carbon allotrope features a flat multi-ring structure that balances mechanical stability and efficient ion transport. The multi-ringed structure DOTT-C creates a distinctive porous topology that facilitates rapid lithium-ion diffusion and stable adsorption sites. These attributes position DOTT-Carbon as a promising high-performance anode material for LIBs. 

\section{Methodology}

The structural, mechanical, and lithium-ion adsorption properties of DOTT-C were investigated using DFT calculations within the CASTEP code \cite{clark2005first}. All calculations employed the Perdew-Burke-Ernzerhof (PBE) exchange-correlation function under the generalized gradient approximation (GGA) scheme \cite{perdew1996generalized}. A plane-wave energy cutoff of 450 eV and a Monkhorst-Pack k-point mesh of $10\times 10\times 1$ were adopted for Brillouin zone sampling. Convergence criteria for energy and forces were set to $1.0 \times 10^{-5}$ eV and $1.0 \times 10^{-3}$ eV/\r{A}, respectively. Structural relaxations were performed with periodic boundary conditions, ensuring minimal residual forces and pressure below 0.01 GPa. A vacuum space of 20 \r{A} was applied in the out-of-plane direction to eliminate interlayer interactions.

Phonon calculations used density functional perturbation theory (DFPT) \cite{baroni2001phonons}. AIMD simulations were carried out at 1000 K for 5 ps with a timestep of 1 fs to assess thermal stability. These simulations employed the Nosé-Hoover thermostat \cite{nose1984unified} within the NVT ensemble, monitoring structural integrity to detect any potential bond-breaking or atomic reconfigurations. The electronic band structure and density of states (DOS) were analyzed using a $k$-point grid of $10\times10\times1$ and $20\times20\times1$, respectively. Optical properties were further explored by calculating the complex dielectric constant under an external electric field, following established methodologies \cite{lima2023dft}.

Lithium adsorption studies were conducted by placing Li atoms at various adsorption sites on the DOTT-C surface, followed by geometry optimization to determine the most energetically favorable configurations. The adsorption energy ($E_{\text{ads}}$) was calculated as $E_{\text{ads}} = E_{\text{Li+DOTT-C}} - \left( E_{\text{DOTT-C}} + E_{\text{Li}} \right)$, where $E_{\text{Li+DOTT-C}}$ is the total energy of DOTT-C with adsorbed Li, $E_{\text{DOTT-C}}$ is the energy of pristine DOTT-C, and $E_{\text{Li}}$ is the energy of an isolated Li atom. Lithium diffusion barriers were determined using the nudged elastic band (NEB) method within CASTEP \cite{makri2019preconditioning,barzilai1988two,bitzek2006structural}. The diffusion coefficient was estimated using the adsorption energies to quantify lithium-ion mobility across the DOTT-C surface.

The mechanical properties of DOTT-C were explored through classical reactive molecular dynamics simulations using the Large-scale Atomic/Molecular Massively Parallel Simulator (LAMMPS) \cite{plimpton1995fast}. A reactive force field derived from a moment tensor potential (MTP) \cite{novikov2020mlip,shapeev2020elinvar} was employed and developed with assistance from MLIP package \cite{mortazavi2021first,podryabinkin2017active}. Training datasets for the MTP were constructed from stress-free and uniaxially strained supercells at varying temperatures, employing AIMD simulations within the Vienna Ab initio Simulation Package (VASP) \cite{kresse1993ab,kresse1996efficiency,kresse1994norm}. Newton's equations of motion were integrated using the velocity Verlet algorithm with a timestep of 0.05 fs. Classical MD simulations were thermalized in the NPT ensemble for 100 ps. Elastic properties and fracture behavior using MLIPs were assessed at 300 K under uniaxial tensile strain with a constant engineering strain rate of $10^{-6}$ fs$^{-1}$. Using MLIP to investigate the mechanical properties of nanomaterials has consistently demonstrated its effectiveness and reliability \cite{mortazavi2023machine,mortazavi2025exploring,mortazavi2023atomistic,mortazavi2020efficient,mortazavi2023electronic,mortazavi2024goldene}.

\section{Results}

\begin{figure}[!htb]
    \centering
    \includegraphics[width=0.8\linewidth]{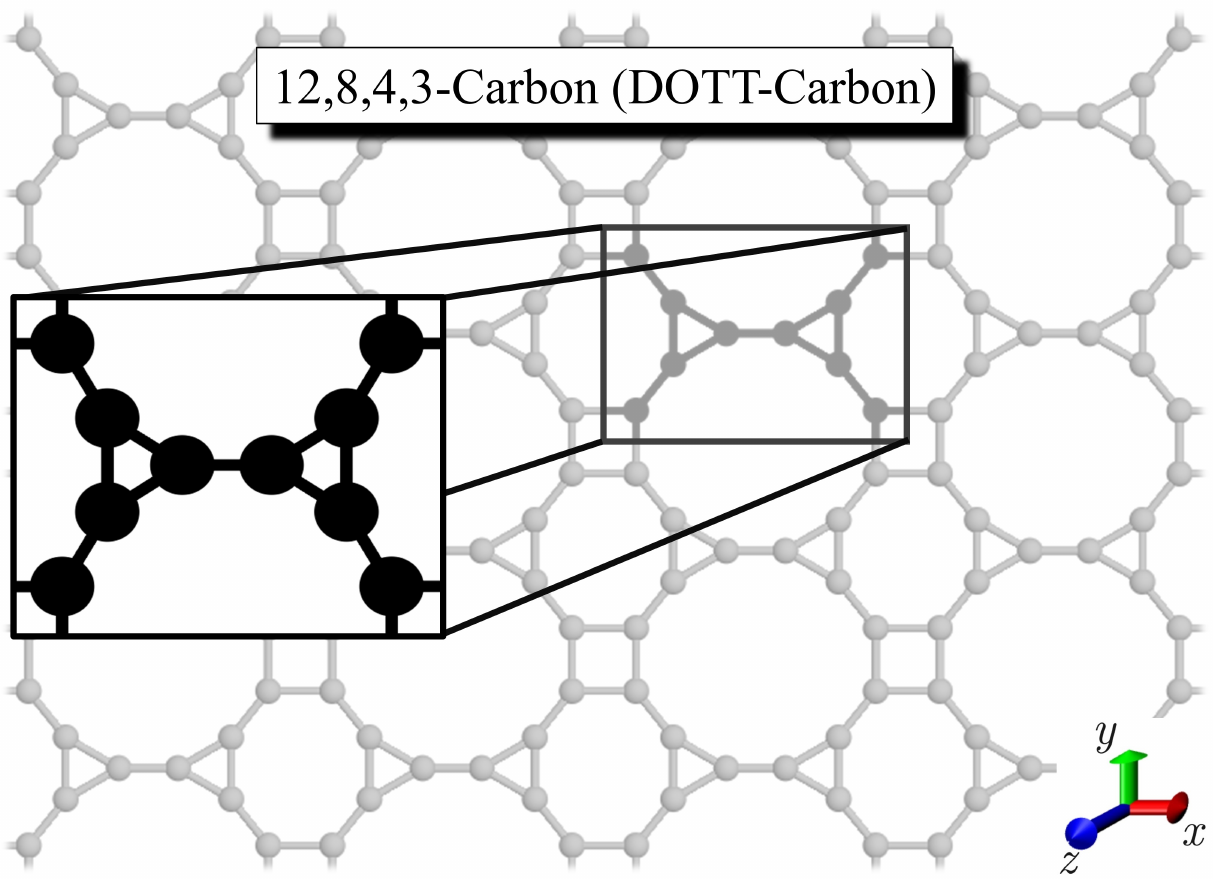}
    \caption{Atomic structure of DOTT-C. The unit cell (in black) has dimensions of $a=9.46$ \r{A} ($x$-direction) e $b=6.08$ \r{A} ($y$-direction).}
    \label{fig:system}
\end{figure}

Figure \ref{fig:system} illustrates the atomic structure of the DOTT-C lattice. The unit cell, highlighted in black, consists of 10 carbon atoms and has lattice parameters of $a=6.58$ \r{A} along the x-direction and $b=5.66$ \r{A} along the y-direction. The C--C bond lengths exhibit slight variations, ranging from 1.37 \r{A} to 1.48 \r{A}. These bond lengths align with expected values for sp$^{2}$-hybridized carbon, indicating strong covalent atomic interactions within the lattice. DOTT-C belongs to the P2/M space group and adopts C2H-1 symmetry, reflecting its symmetric arrangement.

DOTT-C presents a formation energy of -8.25 eV/atom, indicating strong energetic stability. This value is comparable to other 2D carbon allotropes, such as graphene (-8.83 eV/atom) \cite{skowron2015energetics}, sun-graphyne (-8.57 eV/atom) \cite{tromer2023mechanical}, and dodecanophene (-8.19 eV/atom) \cite{lima2024dodecanophene}, graphenylene (-8.28 eV/atom) \cite{zhang2023thermal}, porous-graphene (-6.93 eV/atom) \cite{laranjeira2024graphenyldiene}, T-graphene (-7.87 eV/atom) \cite{majidi2017density}, biphenylene (-7.95 eV/atom) \cite{chen2023biphenylene}, Graphenyldiene (-7.30 eV/atom) \cite{laranjeira2024graphenyldiene}, and Irida-graphene (-6.96 eV/atom) \cite{junior2023irida}. While slightly lower than graphene, it is significantly more stable than several porous counterparts, including graphdiyne (-0.77 eV/atom) and $\gamma$-graphyne (-0.92 eV/atom) \cite{zhao2013two}. 

\begin{figure}[!htb]
    \centering
    \includegraphics[width=0.6\linewidth]{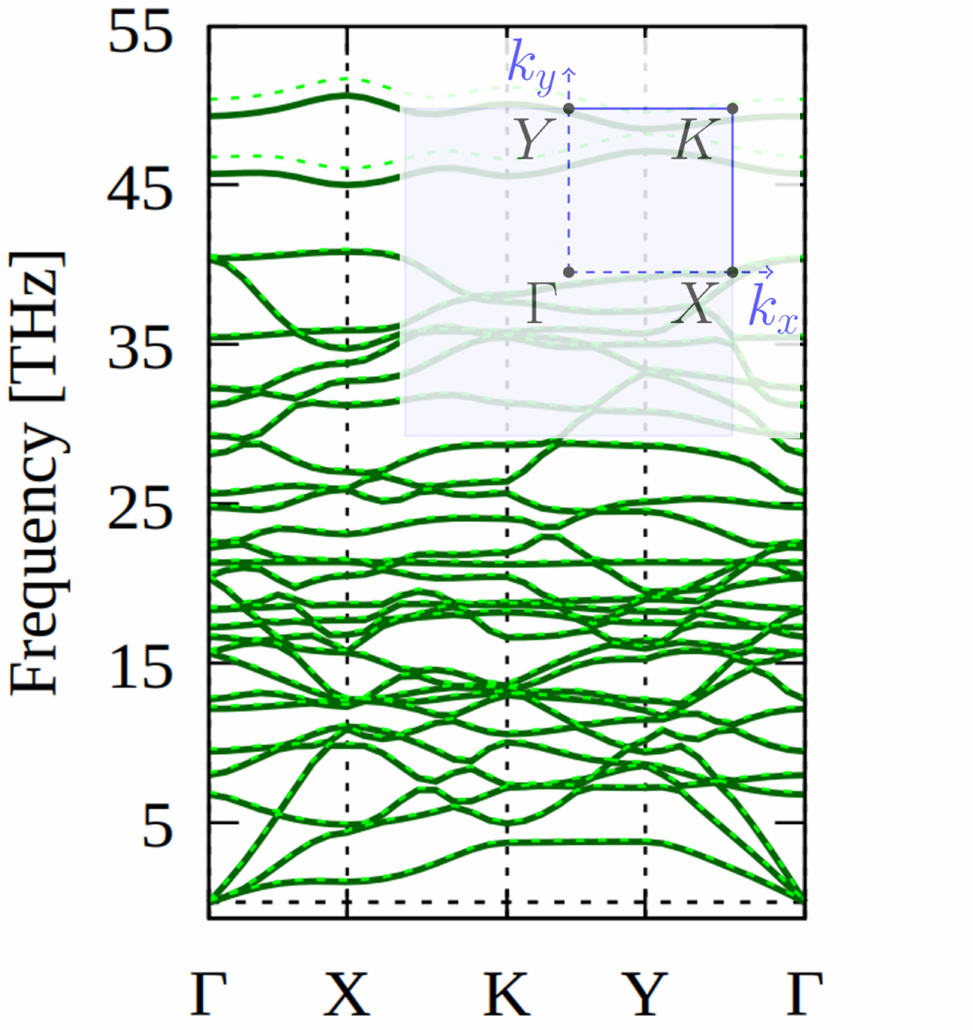}
    \caption{Phonon dispersion analysis of DOTT-C calculated using DFPT (dark green) and MTP (light green) methods. The inset highlights the first Brillouin zone.}
    \label{fig:phonons}
\end{figure}

Phonon dispersion results of DOTT-C, obtained using DFPT (dark green curves) and MTP (dotted light green curves), are shown in Figure \ref{fig:phonons}. The absence of imaginary frequencies points to its good dynamical stability, demonstrating that the structure can remain mechanically stable under moderate perturbations. The phonon branches exhibit smooth and continuous trends, characteristic of well-relaxed crystal lattices. The agreement between MTP-based results and DFPT outcomes validates the accuracy of the generated MLIP specifically for DOTT-C. This alignment also supports the application of ML-based potentials in classical reactive MD simulations conducted in this study. Moreover, the consistency between these methods enables efficient simulations of larger systems and longer timescales without compromising accuracy.

MLIP-based reactive MD simulations are crucial for studying the mechanical properties of nanostructured materials once they combine quantum-level accuracy with the computational efficiency of classical simulations. Unlike purely ab initio approaches, which are computationally expensive for large systems and long timescales, MLIP models are trained on high-fidelity data from DFT calculations, enabling precise predictions of interatomic interactions. Importantly, reactive MD simulations can also capture the formation and breaking of chemical bonds --- a critical feature for understanding fracture mechanics and deformation in nanomaterials --- while maintaining computational costs comparable to conventional classical MD simulations. This feature makes them adequate for exploring complex mechanical behaviors, such as anisotropic stress responses and strain-induced bond reconfigurations, in systems with intricate atomic arrangements. Therefore, the validated MLIP potential for DOTT-C allows for accurate modeling of its multi-ringed structure under mechanical stress, providing deep insights into its fracture dynamics and mechanical resilience.

\begin{figure}[!htb]
    \centering
    \includegraphics[width=0.7\linewidth]{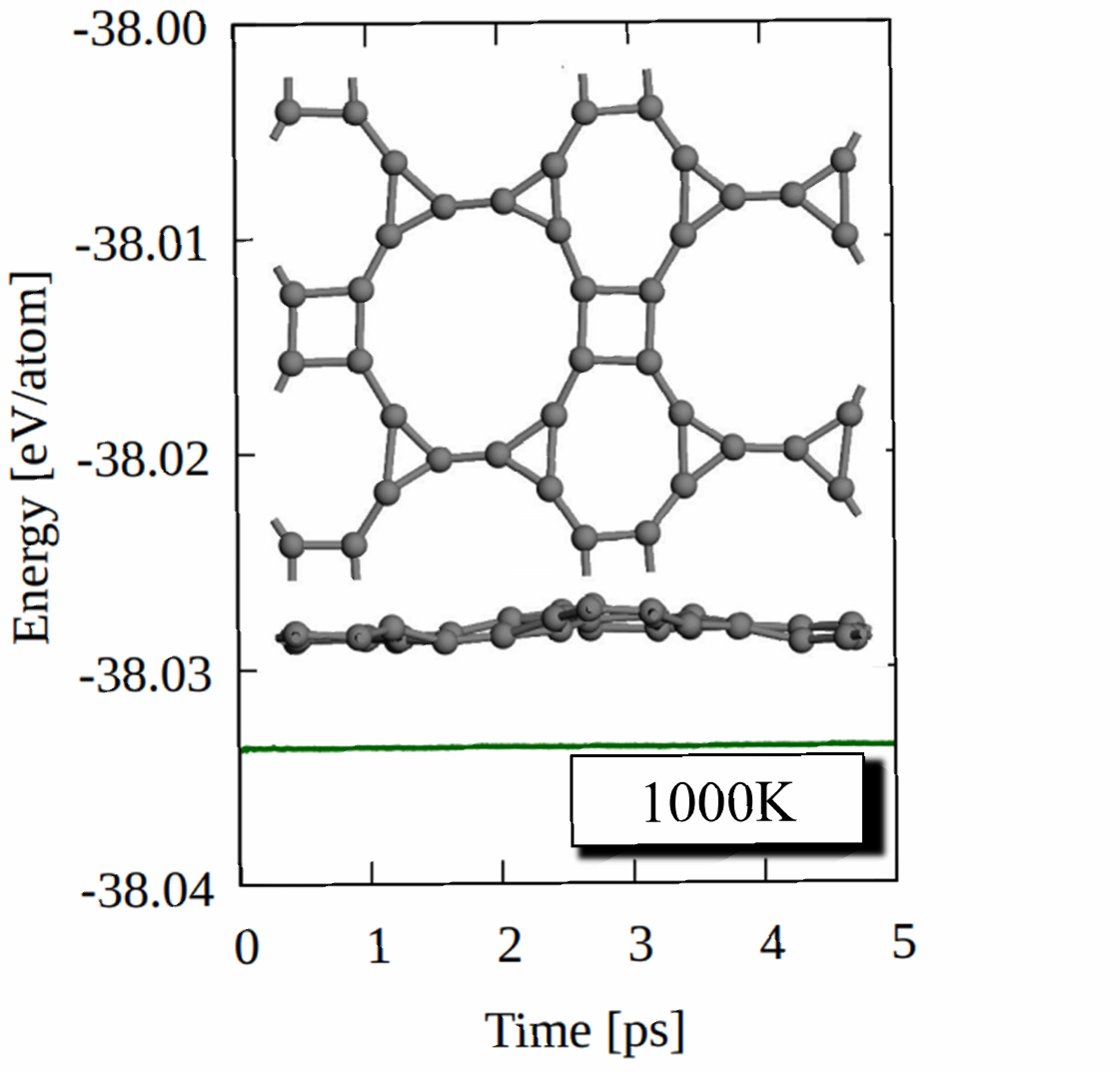}
    \caption{The AIMD simulation results at 1000 K for the lattice total energy as a function of time. Insets show top and side views of the DOTT-C structure for the final simulation stage.}
    \label{fig:aimd}
\end{figure}

Figure \ref{fig:aimd} illustrates the thermal stability of DOTT-C as assessed through AIMD simulations at 1000 K. The insets depict the top and side views of the structure from the final AIMD snapshot at 5 ps. No bond breaking or significant structural reconfigurations are observed during the simulation, confirming the thermal resilience of DOTT-C under elevated temperature conditions. The absence of substantial in-plane and out-of-plane distortions suggests its practical use in nanoelectronic applications operating at high temperatures. This feature is interesting for energy storage systems, such as LIB anodes, where materials are often subjected to thermal stress during charge and discharge cycles. 

The flat pattern observed in the energy curve over the time evolution indicates that DOTT-C can keep its stability under extreme conditions. This stability trend arises from its robust bonding framework, which resists thermal fluctuations, preventing atomic rearrangements. Such behavior underscores the thermal resilience of DOTT-C, which is a critical feature for applications like LIB anodes, where materials undergo repetitive thermal cycling during operation.  

\begin{figure*}[!htb]
    \centering
    \includegraphics[width=\linewidth]{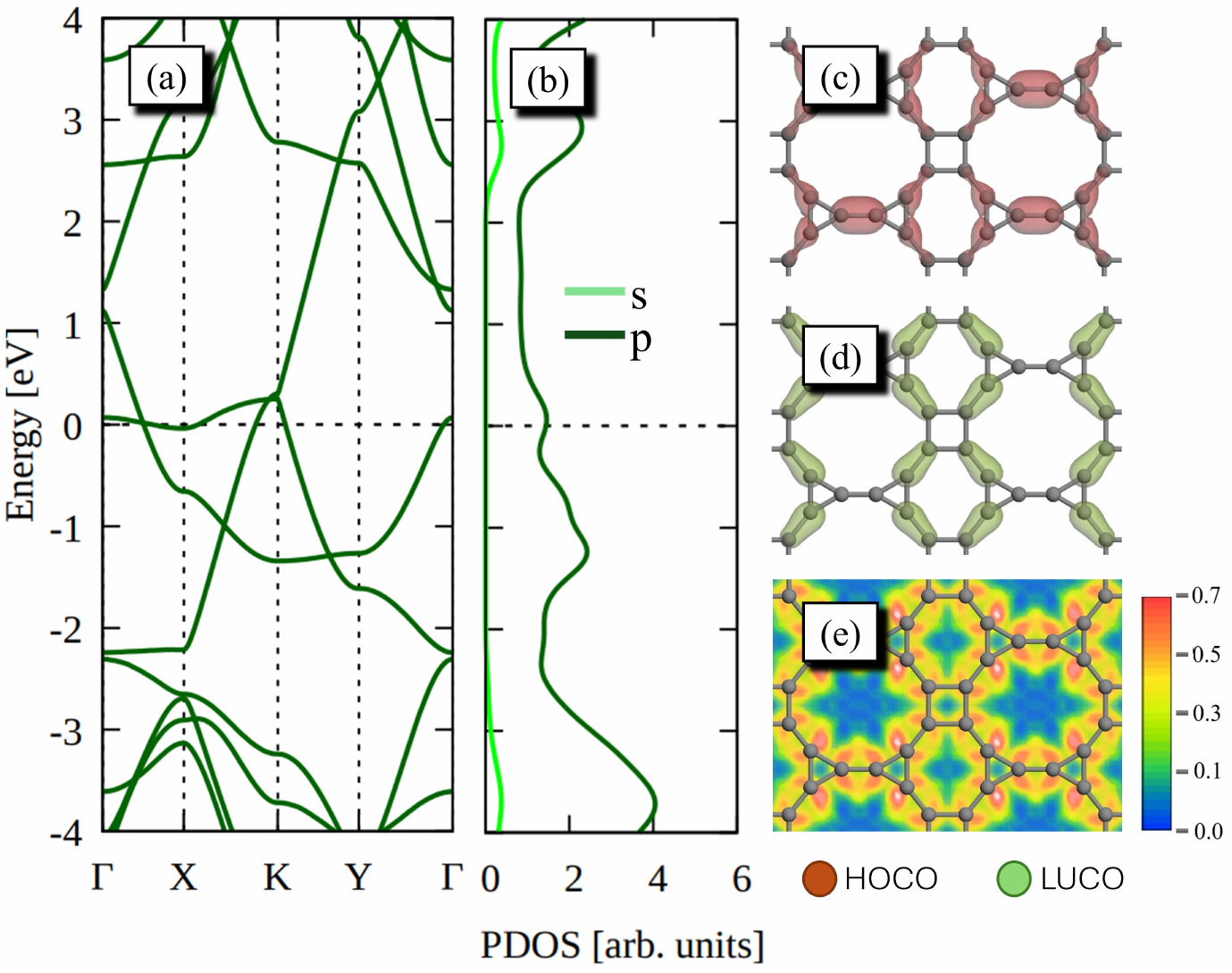}
    \caption{Electronic structure of DOTT-C. (a) Band structure along high-symmetry $k$-paths and (b) PDOS. Panels (c) and (d) depict the Highest Occupied Crystalline Orbital (HOCO, red) and Lowest Unoccupied Crystalline Orbital (LUCO, green) distributions. Panel (e) illustrates the electron localization function (ELF).}
    \label{fig:bands-orbitals}
\end{figure*}

We now explore the electronic properties of DOTT-C, highlighting its potential for efficient charge transport and energy storage applications. Figure \ref{fig:bands-orbitals}(a) presents the electronic band structure of pristine DOTT-C along high-symmetry k-paths, as shown in Figure \ref{fig:phonons}. One can note that the material exhibits a metallic nature with bands crossing the Fermi level. This metallic behavior suggests that DOTT-C can efficiently conduct electrons, a critical property for its application as an anode material in lithium-ion batteries. Figure \ref{fig:bands-orbitals}(b) shows the projected density of states (PDOS), where the contributions from s- (light green) and p-orbitals (dark green) are depicted. The dominant p-orbital contributions near the Fermi level points the role played by carbon's sp$^2$-hybridized electrons in dictating the material's electronic properties.

Figures \ref{fig:bands-orbitals}(c) and \ref{fig:bands-orbitals}(d) illustrate the spatial distribution of the highest occupied crystalline orbital (HOCO, red) and the lowest unoccupied crystalline orbital (LUCO, green), respectively. The HOCO is delocalized across the lattice's 3- and 8-membered rings, while the LUCO concentrates primarily on 4- and 12-membered rings. These orbital distributions show distinct channels for electron and hole transport. In this way, DOTT-C can have a certain level of control over charge transport and storage processes. The spatial distribution of HOCO and LUCO also reveals the potential of DOTT-C in minimizing charge recombination, which is advantageous for energy applications.

The electron localization function (ELF) profile of the DOTT-C lattice is depicted in Figure \ref{fig:bands-orbitals}(e), providing insights into the material's bond and electronic localization characteristics. The ELF map points to highly localized electron density on the C-C bonds (red spots), corresponding to the material's strong covalent interactions. The delocalized (yellow) regions observed in larger rings reflect the metallic nature of DOTT-C, facilitating charge transport across the lattice. These results collectively highlight DOTT-C as a candidate for applications requiring an efficient charge transport mechanism.

The color scheme in the ELF plot ranges from blue (0.0) to red (0.7), representing the probability of electron localization within the lattice. Blue regions (low ELF values) correspond to areas of low electron localization, indicating delocalized electronic states. Red regions (high ELF values) represent highly localized electrons, typically associated with strong covalent bonding. The yellow-green intermediate regions indicate partial localization, reflecting the shared nature of electrons in the larger carbon rings.

In the context of DOTT-C, the red spots along the C--C bonds confirm the presence of robust covalent interactions, which are responsible for maintaining the lattice's structural integrity. Simultaneously, the delocalized yellow regions within the larger rings further emphasize DOTT-C's metallic behavior, enabling efficient charge transport pathways. This combination of localized bonding and delocalized electronic states indicates the material's potential for applications requiring high electrical conductivity and mechanical stability, such as energy storage and electronic devices.

\begin{figure*}[!htb]
    \centering
    \includegraphics[width=0.8\linewidth]{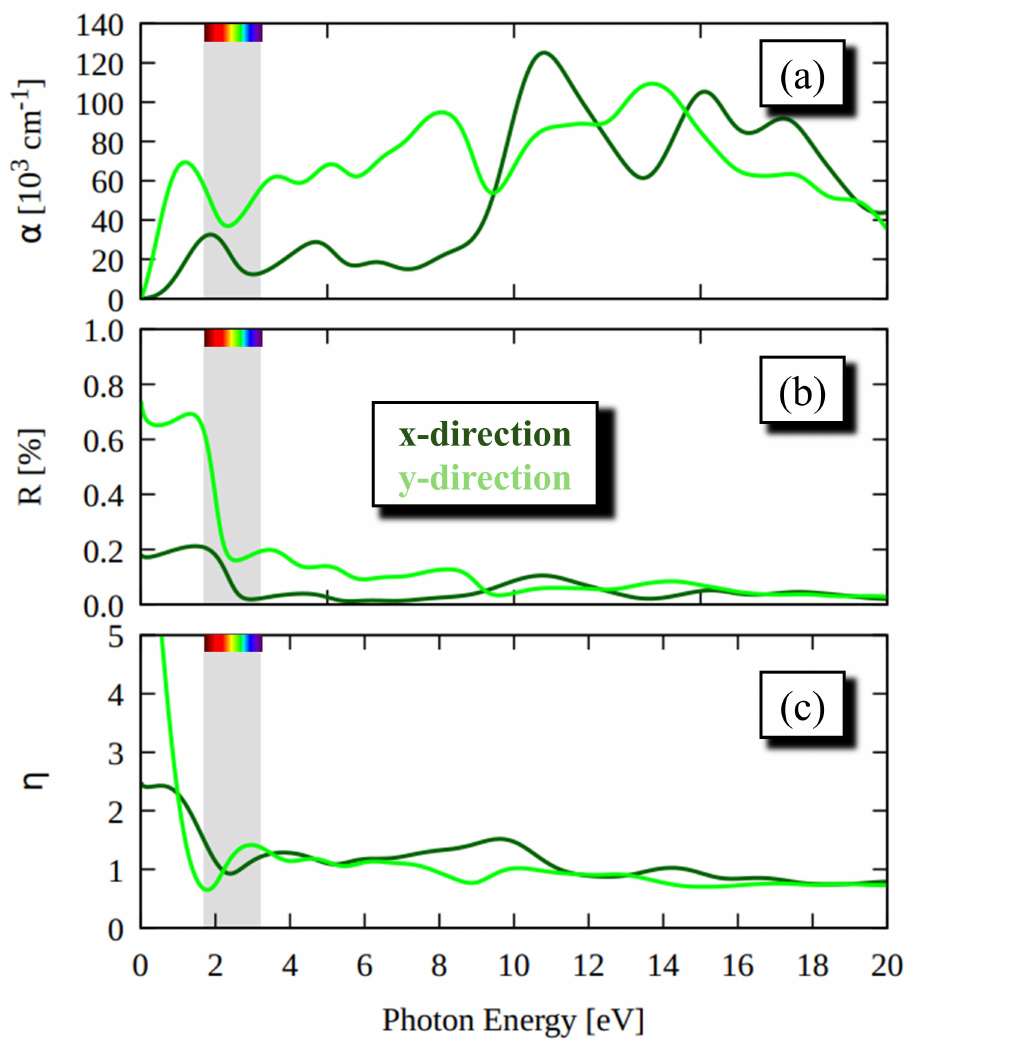}
    \caption{Optical properties of DOTT-C as a function of photon energy and for light polarized along the $x$-direction (dark green) and $y$-direction (light green). Panels (a,b,c) illustrate the results for absorption coefficient ($\alpha$), (b) reflectivity ($R$), and (c) refractive index ($\eta$), respectively. The gray region stands for the visible range of the spectrum.}
    \label{fig:optical}
\end{figure*}

Figure \ref{fig:optical} shows the optical properties of DOTT-C, highlighting its anisotropic behavior along the x- (dark green) and y-directions (light green). In Figure \ref{fig:optical}(a), the absorption coefficient ($\alpha$) is plotted against photon energy, demonstrating broad optical responses spanning from the visible to ultraviolet (UV) range. Several peaks in the absorption spectrum are present, indicating DOTT-C's good absorption efficiency, particularly in the UV range. 

The anisotropic behavior observed in the absorption coefficient ($\alpha$) of DOTT-C, as shown in Figure \ref{fig:optical}(a), stems from its unique multi-ringed lattice structure and the direction-dependent interaction of electromagnetic waves with its electronic states. The distinct orientations of C--C bonds and ring configurations in the x- and y-directions result in varying electronic transition probabilities, leading to distinct absorption profiles for each polarization. Moreover, the intense absorption activity in the UV range is attributed to the delocalized $\pi$-electrons and the material's metallic nature. These features enable high-energy electronic transitions, particularly from bonding to anti-bonding orbitals, which are prevalent in this energy range. As shown above, the dense electronic states near the Fermi level, combined with the strong $\pi$-$\pi^{*}$ transitions, amplify DOTT-C's absorption efficiency in the UV spectrum. This property makes DOTT-C a promising candidate for applications requiring efficient light absorption, such as UV detectors, photovoltaic devices, and other optoelectronic technologies.

The reflectivity ($R$) results are presented in Figure \ref{fig:optical}(b). One can note a low reflectivity across the energy range, consistently below 0.8. The $R$ values are higher within the infra-red range. These values drop dramatically for photon energies higher than 1.9 eV. This behavior suggests that DOTT-C can transmit a significant portion of the incident light and is valid for optical coatings or light-filtering applications. The anisotropic nature of reflectivity is more pronounced in the infrared region. The pronounced anisotropy in reflectivity ($R$) within the infrared range, as shown in Figure \ref{fig:optical}(b), arises from the direction-dependent electronic structure and bonding configuration of DOTT-C. The material's multi-ringed lattice creates an uneven distribution of electronic states, which interacts differently with incoming electromagnetic waves depending on the polarization. In the infrared range, this anisotropy is amplified due to the involvement of low-energy vibrational modes and charge polarization effects, which vary between the x- and y-directions, leading to distinct reflectivity values.

It is worth mentioning that the abrupt drop in reflectivity for photon energies above 1.9 eV and the tendency toward zero reflectivity beyond the visible range is due to the material's strong absorption in the UV region, as observed in panel (a). At these higher photon energies, the incident light is predominantly absorbed rather than reflected, owing to the high density of electronic transitions facilitated by the delocalized $\pi$-electrons in the lattice. This sharp transition to near-zero reflectivity highlights DOTT-C's efficiency in absorbing high-energy photons, making it highly suitable for applications in optical coatings and filters where minimal reflection and maximal transmission of specific light wavelengths are desired.

In Figure \ref{fig:optical}(c), the refractive index ($\eta$) is displayed as a function of photon energy, also showing polarization-dependent behavior. The refractive index decreases with increasing photon energy and converges to approximately 1.0 at higher energies. This trend highlights the birefringent properties of DOTT-C, where $\eta$ varies with light polarization, pointing to its applicability in devices that rely on controlled light propagation, such as waveguides and optical sensors. The rapid drop and eventual convergence of the refractive index ($\eta$) to approximately 1.0 at higher photon energies, as seen in Figure \ref{fig:optical}(c), can be explained by the material's electronic response to electromagnetic waves. At lower photon energies, the interaction of light with the electronic cloud of DOTT-C is more significant, leading to higher refractive indices. However, as the photon energy increases, the electrons in the lattice can no longer respond effectively to the rapidly oscillating electric field of the incoming light. This diminished interaction causes $\eta$ to decrease sharply.

Another relevant result is that the convergence of $\eta$ to 1.0 at high photon energies indicates that the material becomes optically transparent in this range as light passes through without significant refraction or reflection. This behavior is characteristic of many materials where the electronic transitions occur predominantly in lower energy ranges, leaving the higher-energy photons less affected. The anisotropic nature of $\eta$, evident in its polarization-dependent behavior, further emphasizes DOTT-C's potential in birefringent applications, such as waveguides, polarizers, and optical sensors, where precise control of light propagation and polarization is crucial.

\begin{figure*}[!htb]
    \centering
    \includegraphics[width=0.7\linewidth]{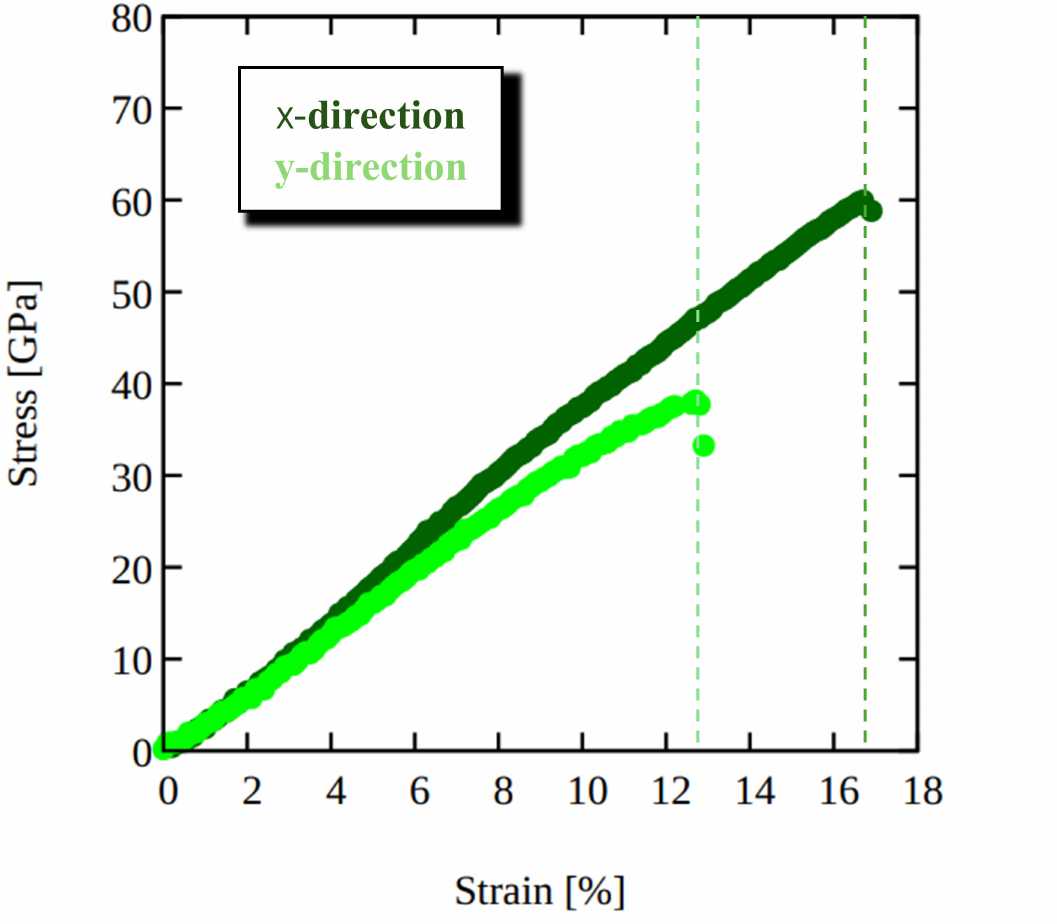}
    \caption{Stress-strain curve of DOTT-C under uniaxial stress in the $x$-direction (dark green) and $y$-direction (light green), calculated using the MLIP method.}
    \label{fig:mechprop}
\end{figure*}

The mechanical response of DOTT-C under uniaxial stress, evaluated using the MLIP method, is shown in Figure \ref{fig:mechprop}. The stress-strain curves are plotted for the x-direction (dark green) and y-direction (light green), showcasing the anisotropic mechanical properties of DOTT-C under tensile loading. The linear region of the curves at lower strains corresponds to the elastic behavior of the material. At higher strain values, a peak is observed, indicating the ultimate stress (maximum stress) DOTT-C can sustain before fracturing.

In the x-direction, DOTT-C achieves an ultimate stress of approximately 60 GPa at a fracture strain of 16.9\% (dashed dark green line). In contrast, the y-direction displays a lower ultimate stress of around 39 GPa, with a fracture strain of 12.8\% (dashed light green line). These distinct responses in the two directions reveal the anisotropic nature of DOTT-C, which arises from the alignment of its multi-ringed structure. The variation in ring configurations between the x- and y-directions influences the material's bond stretching and deformation under applied stress.

From these stress-strain curves, Young's modulus of DOTT-C is determined to be approximately 331.75 GPa in the x-direction and 281.55 GPa in the y-direction, further highlighting its anisotropic mechanical properties. While DOTT-C exhibits significant stiffness, its Young's modulus is lower than that of graphene (near 1 TPa) \cite{scarpa2009effective,lee2012estimation}, which can be attributed to the porous, multi-ring structure of DOTT-C. While imparting rigidity, this architecture introduces additional flexibility compared to graphene's dense hexagonal framework. The higher stiffness and ultimate stress observed along the x-direction reflect the structural alignment and bonding configuration of DOTT-C, enabling it to resist deformation more effectively in that orientation.

\begin{figure*}
    \centering
    \includegraphics[width=\linewidth]{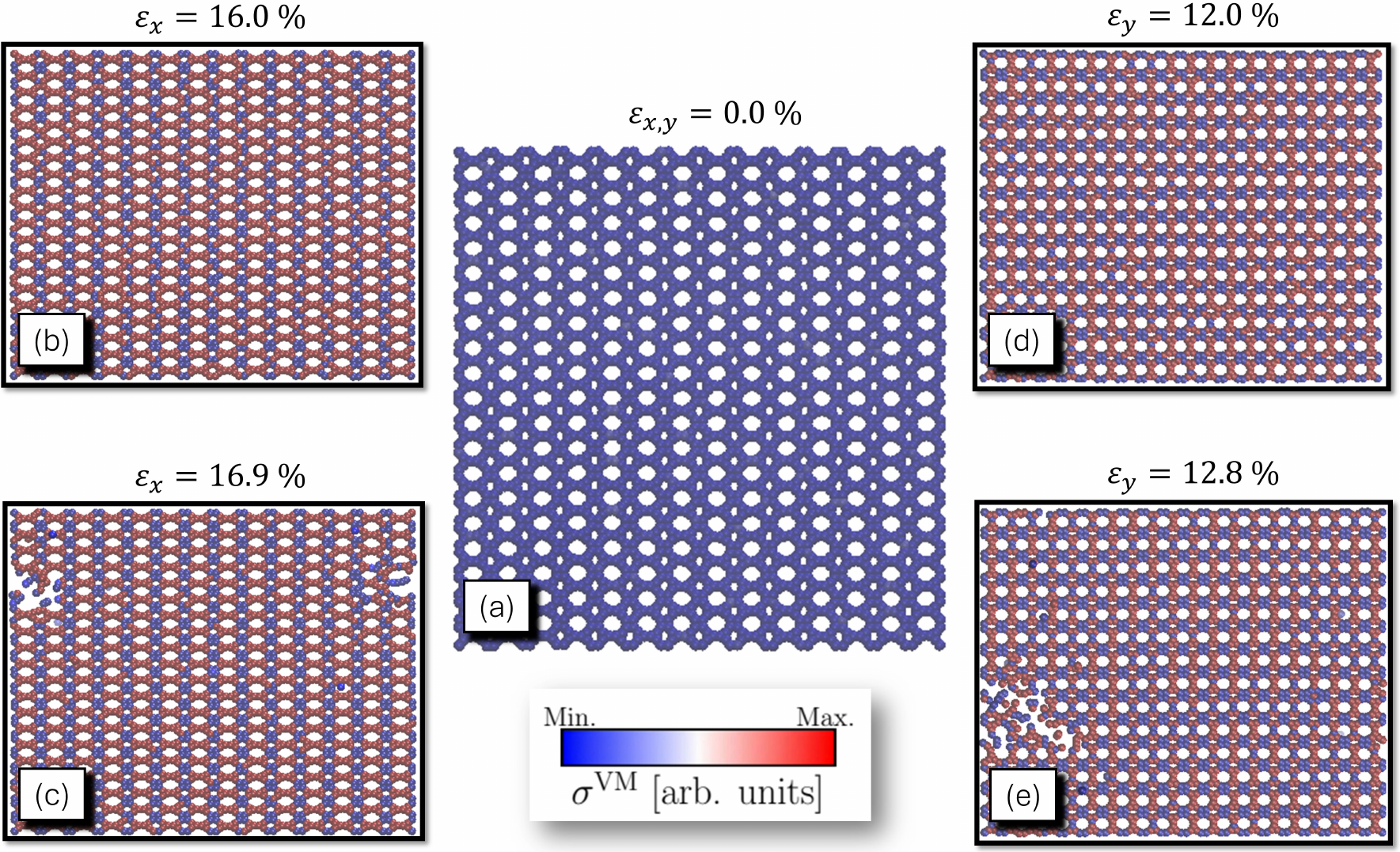}
    \caption{Fracture patterns and stress distribution in DOTT-C, initially unstressed in panel (a), under uniaxial strain in the $x$-direction (panels b and c) and $y$-direction (panels d and e). The von Mises stress distribution is illustrated with a color gradient, where blue represents low stress and red indicates high stress.}
    \label{fig:snapshots}
\end{figure*}

Figure \ref{fig:snapshots} illustrates the fracture patterns and stress distribution in DOTT-C under uniaxial strain applied along the x- and y-directions. The von Mises stress ($\sigma^{\text{VM}}$) distribution, represented by a color gradient ranging from blue (low stress) to red (high stress), provides insights into how strain impacts different regions of the lattice at various deformation stages and highlights the material's fracture behavior.

Figure \ref{fig:snapshots}(a) shows the unstressed DOTT-C lattice without visible stress concentrations. Under strain in the x-direction (panels b and c), stress concentrations form around the multi-ring junctions and larger rings. No bond ruptures occur at a strain of 16.5\% (panel b) at these high-stress regions. By 16.9\% strain (panel c), micro-cracks propagate parallel to the loading direction, eventually coalescing into more extensive fractures, leading to significant lattice rupture. This behavior indicates that the structural alignment of the larger rings along the x-direction introduces weak points, guiding the fracture pathways through these regions.

When strain is applied along the y-direction (panels d and e), the fracture dynamics shift due to the differing orientation of the rings relative to the applied stress. At 12.6\% strain (panel d), stress accumulates more uniformly across the lattice. By 12.8\% strain (panel e), localized stress concentrations cause rapid crack propagation parallel to the strain direction. These cracks expand into an interconnected fracture network, resulting in complete structural failure.

The observed parallel fracture pattern in DOTT-C under uniaxial strain, as depicted in Figure \ref{fig:snapshots}, highlights a unique mechanical behavior distinct from many other 2D carbon allotropes \cite{peng2024molecular,peng2025atomistic,peng2024mechanical,xiao2024effects,junior2022thermal,pereira2022mechanical}. This behavior can be attributed to the structural arrangement of the multi-ringed lattice. In DOTT-C, the larger 12-membered rings and multi-ring junctions align along the direction of applied strain, acting as stress concentrators. These structural features serve as inherent weak points where bonds are more prone to rupture under tension, guiding the fracture pathways along the strain direction. In contrast, other 2D carbon allotropes, such as graphene \cite{felix2020mechanical} or graphyne \cite{cranford2011mechanical}, often exhibit perpendicular fracture patterns due to their continuous hexagonal or uniformly distributed porous networks. In these materials, stress propagates orthogonally to the applied strain because the uniformly distributed bonds offer similar resistance in all directions, leading to more isotropic crack propagation.

\begin{figure*}
    \centering
    \includegraphics[width=\linewidth]{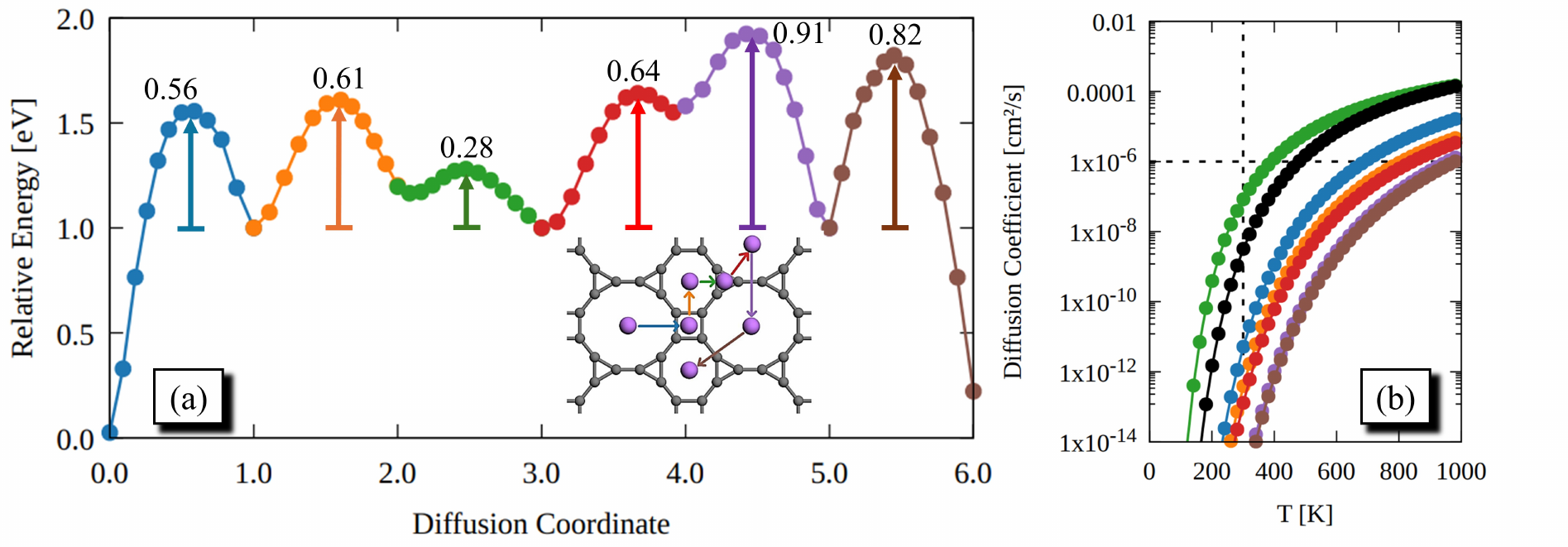}
    \caption{Lithium adsorption and diffusion properties of DOTT-C. Panel (a) illustrates the adsorption energy landscape pathways and profile for a single Li atom migrating on the DOTT-C surface. Panel (b) depicts the temperature-dependent diffusion coefficients. The colored lines in panels refer to the colored arrows in panel (a). For comparison, the black circles refer to the Li-ion diffusion across two DOTT-C sheets (see Figure \ref{fig:2sheets}). The black dotted lines in panel (b) represent the Li-ion mobility for graphene at 300K.}
    \label{fig:lithium}
\end{figure*}

Finally, we present a detailed analysis of the lithium adsorption and diffusion behavior on the DOTT-C surface, highlighting its potential as an anode material for lithium-ion batteries. Figure \ref{fig:lithium}(a) illustrates the energy profile for a single lithium-ion diffusing on the DOTT-C lattice, revealing adsorption sites with varying energy levels. The calculated adsorption energies range between -2.3 eV and -0.89 eV, indicating strong binding between Li ions and specific regions of the lattice. These values compare favorably to other 2D carbon allotropes, such as graphene (-1.57 eV) \cite{kim2014hydrogen}, underscoring DOTT-C's ability to stably adsorb Li ions without compromising structural integrity.

The calculated diffusion energy barriers along different pathways on the DOTT-C surface are highlighted in Figure \ref{fig:lithium}(a). The barriers range between 0.28 and 0.91 eV, among the lowest reported for 2D carbon allotropes. These low barriers highlight the synergy between DOTT-C's porous architecture and electronic structure, facilitating efficient Li-ion transport. The directional dependence of the barriers further emphasizes DOTT-C's anisotropic nature, suggesting that specific pathways can be optimized to enhance ion mobility.

\begin{figure*}
    \centering
    \includegraphics[width=0.6\linewidth]{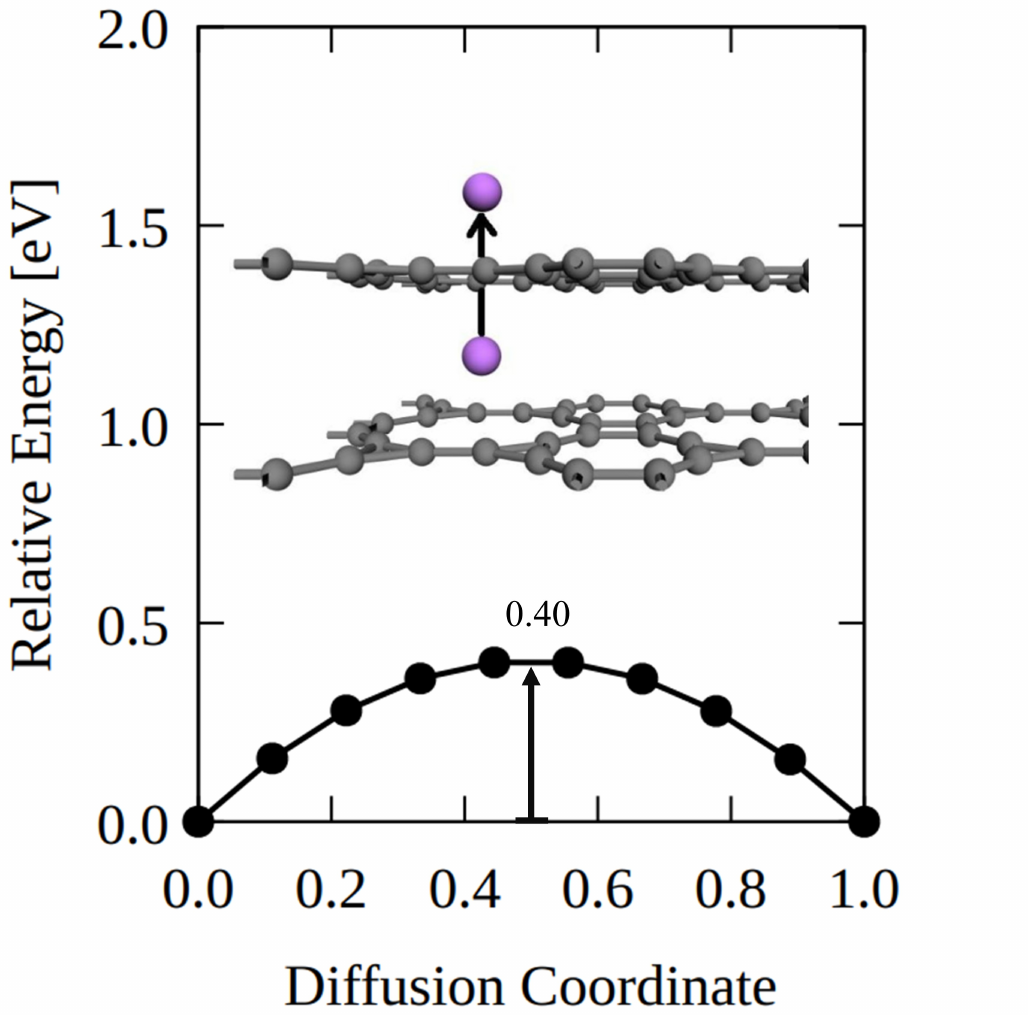}
    \caption{Relative energy profile for a lithium atom diffusion across two DOTT-C sheets. The inset represents the migration pathway.}
    \label{fig:2sheets}
\end{figure*}

Figure \ref{fig:lithium}(b) displays temperature-dependent diffusion coefficients, providing insights into DOTT-C's performance under various thermal conditions. At room temperature, the diffusion coefficient exceeds $1\times 10^{-6}$ cm$^2$/s, demonstrating exceptional Li-ion mobility compared to graphene and other 2D materials. For comparison, the black dotted lines represent the Li-ion mobility for graphene at 300K, and the black circles refer to the Li-ion diffusion across two DOTT-C sheets (see Figure \ref{fig:2sheets}). As temperature increases, the exponential growth in diffusion coefficients validates the thermally activated nature of Li-ion dynamics on DOTT-C's surface. The strong temperature dependence highlights the material's adaptability for high-efficiency lithium-ion batteries operating across various conditions.

The relative energy profile for lithium diffusion across two DOTT-C sheets is presented in Figure \ref{fig:2sheets}. The energy profile highlights a diffusion barrier of 0.40 eV, a critical parameter for evaluating the ease of lithium-ion mobility across the material. The inset shows a schematic representation of the lithium-ion moving across the lattice, emphasizing the flat topology and accessible pathways provided by DOTT-C's unique multi-ringed structure. The moderate diffusion barrier observed here reflects the synergy between DOTT-C's porous architecture and electronic properties. This sufficiently low barrier ensures efficient lithium-ion transport, making DOTT-C a competitive candidate for lithium-ion battery anodes. The smooth and continuous energy landscape suggests minimal resistance to ion movement, which is essential for high-rate performance in battery applications.

Based on these findings, one can realize the DOTT-C's ability to balance strong lithium adsorption with rapid ion diffusion, combining structural stability with efficient ion transport mechanisms. Such attributes are critical for achieving high energy density and excellent cycling stability in next-generation energy storage systems. This diffusion barrier and the material's favorable adsorption and transport properties further validate its potential for advanced energy applications.

\begin{figure*}
    \centering
    \includegraphics[width=0.6\linewidth]{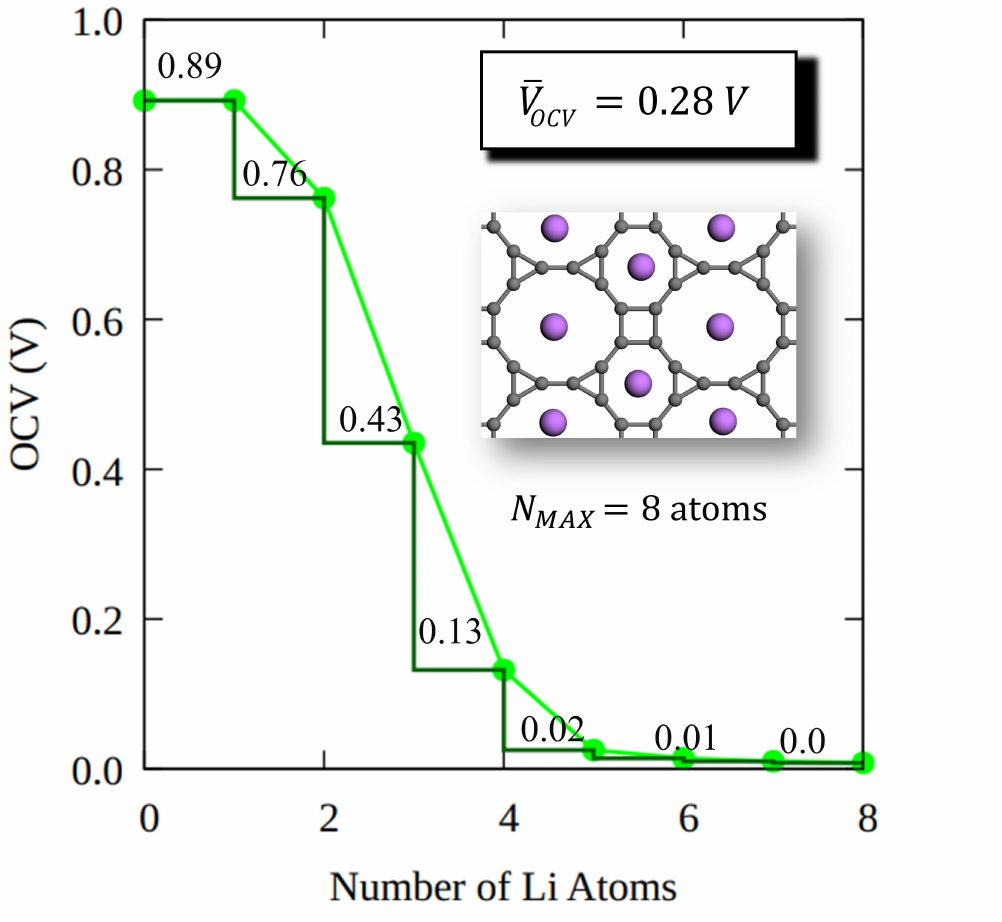}
    \caption{OCV as functions of the number of adsorbed in DOTT-C. The average OCV is 0.28 eV.}
    \label{fig:ocv}
\end{figure*}

Figure \ref{fig:ocv} illustrates the relationship between the open circuit voltage (OCV) and the number of adsorbed lithium atoms on DOTT-C. The OCV values, calculated using a first-principles approach, demonstrate a clear trend as lithium adsorption progresses. At low lithium coverage, the OCV starts high (0.89 V), reflecting strong initial adsorption due to the availability of energetically favorable binding sites. As adsorbed lithium atoms increase, the OCV decreases progressively, stabilizing near 0.0 V after eight adsorbed lithium atoms. This decline in OCV is attributed to increased electrostatic repulsion among the Li ions and competition for adsorption sites as coverage increases.

DOTT-C's average OCV, calculated to be approximately 0.28 V, positions it as a competitive candidate for LIB anode applications. A moderate OCV value is desirable for anodes, as it minimizes the voltage difference between the anode and lithium-ion aggregates, enhancing energy efficiency and reducing the risk of lithium plating during the charging process. For comparison, graphite, a commonly used anode material, has an OCV range of 0.22–0.40 V. At the same time, graphene exhibits an even lower OCV of approximately 0.11 V. These results suggest that DOTT-C offers a balanced OCV that combines stability and efficiency.

The theoretical lithium storage capacity of DOTT-C, calculated as 446.28 mAh/g, exceeds that of graphite (372 mAh/g) \cite{zhang2021graphite} and graphene (568 mAh/g) \cite{liu2013feasibility}, making it a promising alternative for next-generation LIBs. The material's unique multi-ringed porous structure provides ample adsorption sites for lithium ions, ensuring high capacity while maintaining structural integrity. While DOTT-C's capacity is moderate compared to high-capacity materials like twin-graphene (3916 mAh/g), its combination of moderate OCV and robust capacity highlights its potential for practical energy storage applications. The results presented in this Figure \ref{fig:ocv} underscore the effectiveness of DOTT-C's porous architecture in stabilizing Li-ion adsorption while enabling efficient ion transport. The combined metallic nature and low diffusion barriers of DOTT-C point to its potential as a high-performance anode material for lithium-ion batteries, ensuring a balance between energy density, efficiency, and cycling stability.

\section{Conclusions}

This study explored the structural, mechanical, electronic, optical, and lithium-ion storage properties of DOTT-Carbon (DOTT-C), a novel two-dimensional carbon allotrope characterized by a unique multi-ringed porous structure. Using first-principles density functional theory calculations and machine learning interatomic potentials, we demonstrated the potential of DOTT-C as an advanced material for energy storage, particularly as an anode in lithium-ion batteries.

The results revealed that DOTT-C exhibits excellent dynamic and thermal stability, as confirmed by phonon dispersion and ab initio molecular dynamics simulations. Its flat lattice, composed of 12-, 8-, 4-, and 3-membered carbon rings, provides a robust framework capable of withstanding high temperatures and mechanical stress. The stress-strain analysis highlighted its anisotropic mechanical behavior, with Young's modulus values of 331.75 GPa and 281.55 GPa in the xxx- and Y-directions, respectively. The material also demonstrated ultimate tensile strengths of 70 GPa and 40 GPa along these directions, combining stiffness with flexibility due to its porous topology.

DOTT-C is a metallic material with excellent lithium-ion storage properties, with adsorption energies ranging from -2.3 eV to -0.89 eV and a theoretical storage capacity of 446.28 mAh/g. The material's low diffusion barrier of 0.40 eV and high diffusion coefficient of $1 \times 10^{-6}$ cm$^22$/s at room temperature ensures efficient lithium-ion transport, making it well-suited for fast charging and discharging cycles. DOTT-C demonstrated a moderate open circuit voltage (OCV) of 0.28 V, balancing energy efficiency and stability. This moderate OCV reduces the risk of lithium plating during charging and enhances the material's suitability as an anode for LIBs.

\section{Acknowledgements}
\noindent This work received partial support from Brazilian agencies CAPES, CNPq, and FAPDF.
L.A.R.J. acknowledges the financial support from FAP-DF grants 00193.00001808/2022-71 and $00193-00001857/2023-95$, FAPDF-PRONEM grant 00193.00001247/2021-20, PDPG-FAPDF-CAPES Centro-Oeste 00193-00000867/2024-94, and CNPq grants $350176/2022-1$ and $167745/2023-9$. 
L.A.R.J. thanks also to CENAPAD-SP (National High-Performance Center in São Paulo, State University of Campinas -- UNICAMP, proj634) and NACAD (High-Performance Computing Center, Lobo Carneiro Supercomputer, Federal University of Rio de Janeiro -- UFRJ, project a22003) for the computational support provided. The authors acknowledge the National Laboratory for Scientific Computing (LNCC/MCTI, Brazil) for providing HPC resources for the SDumont supercomputer, contributing to the research results reported in this paper.
A.M.A.S. thanks the financial support from Petroliam Nasional Berhad (Petronas).

\bibliographystyle{unsrt}
\bibliography{references}
\end{document}